\documentstyle[preprint,aps]{revtex}
\begin{document}
\draft
\title{Intrinsic Josephson Effect and Violation of the Josephson Relation in
Layered Superconductors}
\author{S.N. Artemenko, A.G. Kobelkov}
\address{
Institute for Radioengineering and Electronics of the Russian Academy of
Sciences, 103907 Moscow, Russia}
\date{\today}
\maketitle
\begin{abstract}
Equations describing the resistive state of a layered superconductor with
anisotropic pairing are derived. The similarity with a stack of Josephson
junctions is found at small voltages only, when current density in the
direction perpendicular to the layers can be interpreted as a sum of the
Josephson superconducting, the Ohmic dissipative and the interference
currents. In the spatially uniform state differential conductivity at higher
voltages becomes negative. Nonuniformity of the current distribution
generates the branch imbalance and violates the Josephson relation between
frequency and voltage.
\end{abstract}

Layered high--$T_c$ superconductors are known to exhibit intrinsic Josephson
effect when the current is flowing across the conducting layers (see {\em
e.g.} \cite{Mue94} and recent measurements \cite{Lat}). Though such materials
can be considered as a stack of 2D--superconductors linked by the Josephson
coupling \cite{LD,Bul}, the theory of the {\em ac} Josephson effect in tunnel
junctions cannot be directly applied to describe the resistive state of
layered superconductors. In a system of series connected junctions the
electric field is located mainly in the insulating barriers due to screening
by the electrons in the metal, and the superconducting banks are in the
equilibrium state. This results in many important consequences including the
Josephson relation between voltage and frequency. But in the layered
materials the superconducting layers are of atomic thicknesses and one must
not ignore nonequilibrium effects which are related to perturbations of the
quasiparticle distribution in the superconductor.  On the other hand, many
experimental evidences for $d$--wave or nearly $d$--wave symmetry of the
superconducting order parameter in layered high--$T_c$ superconductors were
given last years, and a compatibility of the experimental data with $d$--wave
scenario was shown in many theoretical works (see {\em e.g.} \cite{Sc,Ma} and
references therein). In this case the superconducting order parameter has
nodes, {\em i.e.} the quasiparticle density is never exponentially small and
the nonequilibrium effects due to quasiparticles become especially important.
Thus, to understand the intrinsic Josephson effect in high--$T_c$
superconductors one must take into account the nonequilibrium distribution of
the quasiparticles and the relaxation processes in the resistive state.

In this study we calculate current and charge densities in superconductors
with anisotropic pairing as functions of the phase differences of the order
parameter in neighboring layers and of the nonequilibrium scalar potential
related to the quasiparticle branch imbalance, {\em i.e.} the difference
between densities of electron--like and hole--like quasiparticles
\cite{TC,T}. We find that the direct analogy with a stack of Josephson
junctions is limited by nonequilibrium effects and scattering processes, the
difference with Josephson junctions being the most pronounced at lower
temperatures. This results in the negative differential conductivity and in
the violation of the Josephson relation. Similar effects are expected in
layered superconductors with isotropic pairing too, but in the latter case
the quasiparticle density drops down exponentially when temperature decreases
and generation of branch imbalance violating the Josephson relation would be
negligible in the most interesting region of low temperatures.

In our calculations we use the quasiclassical theory of nonequilibrium
superconductivity \cite{LO} modified for the case of layered superconductors
\cite{A80,AK}. We solve the equations in the discrete Wannier representation
for the Keldysh \cite{K} matrix propagator, $\hat G_{nm}$. Its diagonal
components are the retarded and advanced Green's functions, $g^R$ and $g^A$,
and its upper off--diagonal component, $g^K$, is related to the electron
distribution function. We consider the hopping conductivity regime between
the layers, $t_\perp \tau \ll \hbar$, which corresponds to the
case of Josephson interlayer coupling. Here $t_\perp$ is the overlap integral
describing the electron spectrum in perpendicular direction, $\epsilon_\perp
=2t_\perp \cos{dk_\perp}$, $d$ is the lattice constant in the perpendicular
direction, and $\tau$ is the momentum scattering time along the layers. This
approach bears some similarity to the interlayer diffusion model \cite{Gr93}
in which the interlayer coupling is mediated through incoherent hopping
processes, $t_\perp$ being neglected. We assume that a symmetry of the
superconducting order parameter is imposed by the symmetry of the coupling
potential in the selfconsistency condition: thus we do not address the
question of the microscopic nature of the interaction resulting in such a
symmetry. Then the equation for $\hat{G}_{nm}$ has the form
\begin{eqnarray}
-i\hbar\left(\sigma_z \frac{\partial}{\partial t} \hat G_{n\,m} +
\frac{\partial}{\partial t'}\hat G_{n\,m}\sigma_z + {\bf v} \nabla \hat
G_{n\,m}\right)+
t_\perp \sum_{i=\pm 1}(A_{n\,n+i}\hat G_{n+i\,m}-\hat G_{n\,m+i}A_{m+i\,m})+
  \nonumber \\
h_n(t) \hat G_{n\,m} -\hat G_{n\,m} h_m(t')=\frac{i\hbar}{2\tau}
(\langle{\hat G}_{n\,n}\rangle
\hat G_{n\,m }-\hat G_{n\,m}\langle \hat{G}_{m\,m}\rangle)
  \label{G}
\end{eqnarray}
where $\phi$ is the angle of the in-plane electron momentum, $\langle \cdots
\rangle$ means averaging over $\phi$, and $h_n=-i \sigma_y \Delta_n
(\phi)+\mu_n+ \sigma_z {\bf v} {\bf p}_n$. Furthermore, $\mu_n=(\hbar/2)
\partial \chi_n / \partial t +e\Phi_n$ is the gauge invariant scalar potential
in layer $n$, $\chi_n$ is the order parameter phase in the $n$-th layer,
$\Phi_n$ is the electric potential, ${\bf p}_n = (\hbar/2)\nabla \chi_n -
(e/c){\bf A}_n$ is the superconducting momentum parallel to the layers, ${\bf
A}_n$ is the vector potential, and $ A_{n\,m}=\cos{ [(\chi_n-\chi_m)/2] }+
i\sigma_z \sin{[(\chi_n-\chi_m)/2]}$. The Pauli matrices introduced above act
on the spin indices of Green's functions. Each $\hat G_{nm}$ depends on two
times (or on two energies in Fourier representation), and on $\phi$.

The r.h.s. of (\ref{G}) is an elastic-collision integral in Born
approximation. Using Born approximation we neglect low--energy quasiparticle
bound states created by impurities (see \cite{Gr96} and references therein),
and, hence, our results are applicable provided typical energies of
quasiparticles are larger than the bandwidth of the impurity induced bound
states, $T > \sqrt{\hbar T_c/ \tau}$.

We calculate current and charge densities assuming the clean limit, $T_c \tau
\gg \hbar$, and neglect, where possible, the pairbreaking due to elastic
scattering. We assume also the case of frequencies $\hbar\omega$ much smaller
than typical electronic energies: temperature and the amplitude of the gap,
$\Delta_0$.

Consider a current flowing between the layers of a quasi two--dimensional
superconductor, $t_\perp \ll \Delta_0$. When the current exceeds its critical
value it produces nonequilibrium perturbations the scale of which is
determined by the interlayer coupling, and for small $t_\perp$, the value of
the nonequilibrium potential $\mu$ is small as well. Then we may solve
equations (\ref{G}) perturbatively, considering $t_\perp$ and $\mu$ as small
values. We calculate the off--diagonal in layer number components of $g^K$ in
the linear approximation in $t_\perp$, these components are related to the
current density in the direction perpendicular to the layers.  The diagonal
in layer numbers component of Tr$g^K$ determines perturbations of the charge
density, we calculate it up to the second order in $t_\perp$ neglecting $\mu$
in comparison to $\Delta$. The structure of the solution can be demonstrated
by the combination $g_j = A_{n\,n-1}g^K_{n-1\,n}- g^K_{n\,n-1} A_{n-1\,n}$
the integral of which over energies and $\phi$ yields the current density
across the layers \cite{A80}. In the Fourier representation we get
\begin{eqnarray}
g_j= -it_\perp \tanh{\frac{\epsilon}{2T}}\left(\frac{\Delta^2}{(\xi^R)^3}
-\frac{\Delta^2}{(\xi^A)^3}\right)S_\omega
- \sum_{\omega_1}\frac{dn_F}{d\epsilon} \frac{4it_\perp
\theta(|\epsilon|-|\Delta|)}{a(\omega_1 +i\tilde{\nu})}
\Big[\omega_1 \Big(
a^2 s_{\omega_1} c_{\omega-\omega_1}
  \nonumber\\
- c_{\omega_1}s_{\omega-\omega_1} \Big) + \sum_{\omega_2} \frac{\omega_2
(\mu_{n-1}-\mu_n)_{\omega_2}}{\hbar(\omega_2 +i\tilde{\nu}) \langle (\omega_2
+i\nu_b)/ (\omega_2 +i\tilde{\nu}) \rangle} (s_{\omega -\omega_1}
s_{\omega_1-\omega_2} + c_{\omega -\omega_1} c_{\omega_1-\omega_2}) \Big],
  \label{gj}
\end{eqnarray}
where $S$, $s$, and $c$ are Fourier components of $\sin{\varphi_n}$,
$\sin{\varphi_n/2}$, and $\cos{\varphi_n/2}$, respectively, $\varphi_n
=\chi_n-\chi_{n-1}$ is the phase difference between the layers. The first
term in (\ref{gj}) describes perturbations of the retarded and advanced
propagators, it is related to the supercurrent, $\xi^{R(A)} = \pm
\sqrt{(\epsilon \pm i0)^2-|\Delta|^2}$.  The last terms describe
perturbations of the distribution function and contribute to the
quasiparticle current, $n_F$ is Fermi distribution function, and
$a=\epsilon/\xi$, $\xi=\sqrt{\epsilon^2-|\Delta|^2}$.  Furthermore,
$\tilde{\nu} = (1/\tau)[\langle a\rangle - (\Delta/\epsilon)\langle
\Delta/\xi \rangle]$ is effective momentum scattering rate for
quasiparticles. Finally, $\nu_b= (1/ \tau)[\langle a \rangle- a^{-1} -
(\Delta/\epsilon)\langle \Delta/\xi \rangle]$ is the effective branch
imbalance relaxation rate. The branch imbalance is
related to the nonequilibrium potential $\mu$ \cite{TC,T,AV}. In the case of
isotropic pairing elastic scattering does not contribute to the relaxation of
the branch imbalance, and the latter relaxes via energy scattering processes.
We shall consider the opposite case of the order parameter close to
$d$--wave symmetry, when the gap has nodes: $\langle \Delta(\phi) \rangle^2
\ll \langle \Delta(\phi)^2 \rangle$. In this limit we may simplify the
equations because both $\tilde{\nu}$ and $\nu_b$ do not depend on $\phi$,
$\tilde{\nu}=(1/\tau)\langle a \rangle$ and $\nu_b =(1/\tau)\langle \Delta^2
(\phi)a/ \epsilon^2\rangle$. Then the expression for the current density 
between
layers $n$ and $n-1$ acquires the form
\begin{eqnarray}
j_\perp= j_c \sin{\varphi_n} + \frac{1}{ed} \int_{-\infty}^t \frac{dt_1}{\tau}
\Big[ \Big( \int_{-\infty}^{t_1} \frac{dt_2}{\tau} \hat{\sigma}_b(t,t_1,t_2)
[\mu_{n-1}(t_2) -\mu_{n}(t_2)]
  \nonumber \\
+ \hat{\sigma}(t,t_1)\frac{\hbar}{2} \frac{\partial \varphi_n}{\partial
t_1}\Big) \cos{\frac{\varphi_n(t) -\varphi_n(t_1)}{2}} +
\hat{\sigma}_{i}(t,t_1)\frac{\hbar}{2} \frac{\partial \varphi_n}{\partial t_1}
\cos{\frac{\varphi_n(t) +\varphi_n(t_1)}{2}} \Big],
  \label{J}
\end{eqnarray}
where generalized conductivities are given by
\begin{eqnarray}
\hat{\sigma}(t,t_1)=\sigma_{N\perp}\langle \frac{\epsilon}{\xi}
e^{-\tilde{\nu}(t-t_1)} \rangle_\epsilon - \hat{\sigma}_{i}(t,t_1), \;
\hat{\sigma}_{i}(t,t_1)=\sigma_{N\perp} \langle
\frac{\Delta^2}{2\epsilon \xi} e^{-\tilde{\nu}(t-t_1)} \rangle_\epsilon
  \label{s} \\
\hat{\sigma}_b(t,t_1,t_2)=\sigma_{N\perp}\langle \frac{\xi}{\epsilon}
e^{-\tilde{\nu}(t-t_1)}e^{-\nu_b(t_1-t_2)} \rangle_\epsilon
  \label{si}
\end{eqnarray}
Averaging over the angles and quasiparticle energies in (\ref{s}-\ref{si}) is
performed according to $$\langle \cdots \rangle_\epsilon = -
\int_{-\infty}^{\infty} d\epsilon \langle \theta(|\epsilon|-|\Delta(\phi)|)
\frac{dn_F}{d\epsilon}(\cdots ) \rangle.$$
The first term in (\ref{J}) describes the Josephson current, $j_c =
\hbar c^2/(4\pi e\lambda_\perp^2 d)$ being the critical current,
$\lambda_\perp$ is the penetration length for a superconducting current
perpendicular to the layers. The quasiparticle contribution is given by the
last terms containing retardation effects related to the momentum and branch
imbalance relaxation. These effects correspond to factors $(\omega + i \nu)$
in (\ref{gj}).  In the limit of $\Delta \rightarrow 0$ these terms reduce to
the Ohmic current $\sigma_{N\perp}E/ (1+i\omega \tau)$, where
$\sigma_{N\perp}$ is the static conductivity in the normal state. In the low
frequency limit the quasiparticle contribution may be interpreted as the
Ohmic and interference current (the last term), the Ohmic current consisting
of two contributions, one of which is related to the branch imbalance and to
its relaxation. When $\varphi$ is a slowly varying function of time, $\omega
\ll \tilde{\nu}$, the retardation effects can be neglected, and the terms
depending on the phase difference acquire a simple form typical for the
Josephson tunnel junctions:
\begin{equation}
j_\perp= j_c \sin{\varphi_n} + \frac{\hbar}{2ed} \left(
\sigma \frac{\partial \varphi_n}{\partial t}
+ \sigma_{i}\frac{\partial \varphi_n}{\partial t}
\cos{\varphi_n}\right),
  \label{JJ}
\end{equation}
where conductivities are given by (\ref{s},\ref{si}) with the exponents
integrated over time $t_1$, which results in factors $1/\tilde{\nu}$. Near
$T_c$ one gets $\sigma = \sigma_{N\perp}$. At low temperatures, $T \ll
\Delta_0$, the quasiparticle conductivity is of the order of the normal state
conductivity, because the decrease of the normal carrier density upon cooling
is compensated by the decrease of the effective scattering rate of
quasiparticles by the same factor $\propto T/\Delta_0$. For the simplest
angular dependence of the gap parameter with the $d$-wave symmetry, $\Delta
=\Delta_0 \cos{2\phi}$, we find $\sigma = 3\sigma_{N\perp}/4$. For the
interference term we get $\sigma_{i} = (\Delta/T) \,\sigma_{N\perp}$ at $T
\gg \Delta$, and $\sigma_{i} \sim \sigma_{N\perp}/4$ at lower temperatures.

The value of $\mu$ in the current density (\ref{J}) must be determined from
the separate equation describing the branch imbalance dynamics. Such an
equation is given by the Poisson's equation with the charge density
calculated from the integral of ${\bf Tr}\,g^K_{n\,n}$ over energies. In the
Fourier representation we get for the charge density $\rho_n$ in the $n$-th
layer
\begin{equation}
-i\omega \rho_n =-i\omega\gamma \frac{\kappa^2}{4\pi e} \mu_n -
\nabla(\sigma_{2\|} \nabla \mu_n /e + i\omega\sigma_{1\|}{\bf P}_{n}) +
\frac{1}{ed^2} \sum_{\omega_1} (J_n -J_{n-1}),
  \label{mu}
\end{equation}
where $J_n =\sum_{\omega_2} \sigma_{2\perp}(n;\omega, \omega_1,
\omega_2)[\mu_{n}(\omega_2) - \mu_{n-1}(\omega_2)]  -
\sigma_{1\perp}(n;\omega_1, \omega)2i\hbar\omega_1\varphi_n(\omega_1)$
describes the flow of quasiparticles between layers $n$ and $n-1$ generating
the branch imbalance. Note that $\sigma(n)$ in the equations for charge
density depend on the branch imbalance scattering rate, they are different
>from the conductivities for current densities.
\begin{eqnarray}
\gamma=1- \langle \frac{\omega a}{(\omega+i\nu_b)} \rangle_\epsilon,\;
\sigma_{k\|}=\sigma_{N\|}\langle \frac{ia^{1-2k}\omega^k}
{\tau (\omega+ i\tilde{\nu}) (\omega+i\nu_b)^k} \rangle_\epsilon,
  \label{smu}\\
\sigma_{1\perp}(n)=\sigma_{N\perp}
\langle \frac{ia^{-1}\omega(c_{\omega_1} c_{\omega-\omega_1} - s_{\omega_1}
s_{\omega-\omega_1} )}{\tau (\omega_1+ i\tilde{\nu})
(\omega+i\nu_b)}\rangle_\epsilon,
  \label{smu1}\\
\sigma_{2\perp}(n)=\sigma_{N\perp}
\langle \frac{i\omega \omega_2 (a^{-3} c_{\omega_1-\omega_2} 
c_{\omega-\omega_1} -
a^{-1}s_{\omega_1-\omega_2} s_{\omega-\omega_1})}
{\tau (\omega_1+ i\tilde{\nu})(\omega+i\nu_b)(\omega_2+i\nu_b)}
\rangle_\epsilon
  \label{smu2}.
\end{eqnarray}
>From equation (\ref{mu}) one can see that the branch imbalance $\mu_n \neq 0$
is generated by a nonuniform quasiparticle current flow.

In the limit of small phase differences between the layers equations
(\ref{J}-\ref{si}) and (\ref{mu}-\ref{smu2}) describe the linear response
characterized by different conductivities for the response to the solenoidal
and to the potential electric fields \cite{AKd}.

Using the definition of $\mu_n$ we get
\begin{equation}
eV_n -\frac{\hbar}{2} \frac{ \partial \varphi_n}{\partial t} =\mu_n
-\mu_{n-1},
  \label{JR}
\end{equation}
where $V_n \equiv \Phi_n-\Phi_{n-1}$ is the difference of electric potentials
per a layer. Thus, the r.h.s. of (\ref{JR}) describes violation of the
Josephson relation.

We consider, first, the uniform case. In the case of spatially uniform
current distribution both $\nabla {\bf P_n} =0$ and $J_n = J_{n-1}$, and
using in (\ref{mu}) $\rho=0$ we find $\mu_n=0$. Then, according to
(\ref{JR}), the Josephson relation between the frequency and the electric
potential difference per a layer, $V_n$, is satisfied. In the limit of small
frequencies the current--phase relation (\ref{JJ}) has the form similar to
Josephson tunnel junctions, and the effects typical for Josephson junctions
must be observed. However, the analogy is limited by the region of voltages
and frequencies smaller, than the effective momentum scattering rate of the
quasiparticles, $\nu_{qp} \approx \tilde{\nu} (\epsilon=T)$. At higher
voltages the finite scattering time effects, which were neglected in
(\ref{JJ}), destroy the analogy.

In principle, for typical parameters of a superconductor both current biased
and voltage biased regimes in the resistive state are possible. For
simplicity we shall concentrate on the case of the voltage bias, which can be
realized, for example, when a capacity is connected in parallel with the
superconductor. Then we can easily find the time dependence of the phase
difference using the Josephson relation, and calculate the current density
>from equation (\ref{J}).
\begin{equation}
j_\perp = \sigma_{N\perp} \langle \frac{V}{2ad\tau(4\omega^2 +
\tilde{\nu}^2)} [ \tilde{\nu} (1+a^2) +\frac{\Delta^2}{2\xi^2}
\sqrt{4\omega^2 + \tilde{\nu}^2}\cos{\omega t} ] \rangle_\varepsilon,
  \label{IV}
\end{equation}
where $\hbar \omega =2eV$ and we omitted the layer index in the uniform
state. Equation (\ref{IV}) resembles expression for the current density
across a tunnel junction. It contains {\em dc} and {\em ac} components, the
characteristic frequency and voltage of such a junction, $\hbar\omega_c
\equiv 2eV_c$, are determined by the momentum scattering time: $V_c \sim
(\Delta_0/T)^2(\hbar/e\tau)$ at $T \gg \Delta_0$, and $V_c \sim
(T/\Delta_0)(\hbar/e\tau)$ at $T \ll \Delta_0$. At $eV \gg \hbar\nu_{qp}$ the
{\em dc} current decreases with voltage increasing. Note, that at lower
temperatures $eV_c \sim \hbar\nu_{qp}$, thus, the $I-V$ curve starts to
decrease with voltage already at $V \sim V_c$. Though numerically the result
depends on the details of the angle dependence of the order parameter,
qualitatively it can be illustrated by the explicit calculations with $\Delta
=\Delta_0
\cos{2\phi}$ at temperatures $T \ll \Delta_0$.
\begin{equation}
j_\perp = \frac{\sigma_{N\perp}}{8d} \left\{
  \begin{array}{cc}
3V & \mbox{at}\;\;V \ll V_c ,\\
(\pi \hbar T/e\Delta_0 \tau)^2 V^{-1} & \mbox{at}\;\; V \gg V_c.
  \end{array}
\right.
  \label{-}
\end{equation}

The origin of the negative differential conductivity at high voltages can be
interpreted as the decrease of the dissipation at frequencies higher than the
quasiparticle scattering rate, $\omega \gg \nu_{qp}$, or, equivalently, using
the analogy to the negative differential conductivity in semiconductor
superlattices \cite{Tsu} at voltages per period higher, than both the width
of the miniband and the momentum scattering rate. In the latter case the
chemical potential in the adjacent layers is shifted by the value of the
voltage exceeding the width of the band of the allowed electronic states,
therefore, the electron's energy in one layer corresponds to the forbidden
states in the neighboring layer. So, the effect must be present in the normal
state of layered conductors as well (see also \cite{AVFTT}).

The negative differential conductivity indicates to an instability of the
uniform resistive state at voltages $eV > \hbar \nu_{qp}$. The instability
must result in a nonuniform current distribution in which, according to
(\ref{mu}) $\mu\neq 0$ and the Josephson relation is violated. Nonuniform
current distribution may be created also at lower voltages due to many other
reasons {\em e.g.} due to Meisner effect, due to contacts, or due to
nonuniformities of the material. Thus, the correction to the Josephson
relation in (\ref{JR}) depends on the experimental conditions. For
illustration we estimate such a correction for an artificial but easily
treatable model of the nonuniformity created by a layer dependent impurity
scattering time, $\tau_n$. We consider the case when variations of $\tau_n$
with $n$ are small, and the nonequilibrium potential $\mu$ can be calculated
perturbatively. Then using condition $\rho_n=0$ we get for the case of low
frequencies $\omega \ll
\nu_{qp}$
\begin{equation}
\mu_n = \langle \frac{t_\perp^2}{2\hbar a}\left( 
\frac{\dot{\varphi}_n}{\nu_{b\,n}
(\tilde{\nu}_n + \tilde{\nu}_{n-1})} - \frac{\dot{\varphi}_{n+1}}{\nu_{b\,n}
(\tilde{\nu}_n + \tilde{\nu}_{n+1})} \right) \rangle_\epsilon.
  \label{mj}
\end{equation}
Estimating the integrals in (\ref{mj}) for $T \ll \Delta_0$ we get from
(\ref{JR}) $[V_n - (\hbar/2e)\dot{\varphi}_n]/V_n \sim \hbar^{-2} t_\perp^2
\tau_n (\tau_{n+1} - \tau_{n-1})(\Delta_0/T)\ln{\Delta_0\tau}$. One can see
that the significant violation of the Josephson relation may be created even
by small variations of $\tau$ in different layers.

Thus, generally speaking, the Josephson relation in layered superconductors
with Josephson interlayer coupling and anisotropic pairing is satisfied under
the special conditions of the uniform current distribution only, which is
difficult to satisfy. Even in the ideally uniform samples the uniform state
is expected to be unstable at higher voltages because of the negative
differential conductivity. This may be the origin of the irregular character
of the $I-V$ curves observed usually in the measurements of the intrinsic
Josephson effect in high-$T_c$ materials.

We are grateful to ISI Foundation (Torino) where part of this work was done
and supported by EU INTAS Network 1010-CT930055.

\end{document}